\documentstyle[12pt]{article}
\normalbaselines
\begin{document}
\newcommand{\be}{\begin{equation}}
\newcommand{\ee}{\end{equation}}

\begin{center}

{\bf QCD and models on multiplicities in $e^+e^-$ and $p\bar p$ interactions}\\

\vspace{2mm}

I.M. Dremin\footnote{E-mail: dremin@lpi.ru}\\

\vspace{2mm}

Lebedev Physical Institute, Moscow\\

\end{center}

\begin{abstract}
A brief survey of theoretical approaches to description of multiplicity
distributions in high energy processes is given. It is argued that the
multicomponent nature of these processes leads to some peculiar
characteristics observed experimentally. Predictions for 
LHC energies are presented.
It is shown that similarity of the energy dependence of average multiplicities
in different reactions is not enough alone to suggest the universal mechanism of
particle production in strongly-interacting systems. Other characteristics of
multiplicity distributions depend on the nature of colliding partners. 
\end{abstract}

\section{Introduction.}

Multiplicity distributions are the most general characteristics of
any high energy process of multiparticle production. They depend on
the nature of colliding particles and on their energy. Nevertheless, 
it has been found that their shapes possess some common qualitative
features in all reactions studied. At comparatively low energies, these 
distributions are relatively narrow and have sub-Poissonian shapes. 
With energy increase, they widen and fit the Poisson distribution.
At even higher energies, the shapes become super-Poissonian, i.e. 
their widths are larger than for Poisson distribution. The width increases 
with energy and, moreover, some shoulder-like substructures appear. 

Their origin is usually ascribed to multicomponent contents of the process.
In QCD description of $e^+e^-$-processes these could be subjets formed 
inside quark and gluon jets (for the reviews see, e.g., \cite{koch, dgar}).
In phenomenological approaches, the multiplicity distribution in a single 
subjet is sometimes approximated by the negative binomial distribution 
(NBD) first proposed for hadronic reactions in \cite{giov}. For 
hadron-initiated processes, these peculiarities are often ascribed 
to multiple parton-parton collisions \cite{lpol, hump, goul, e735, walk}, 
which could lead, e.g., to two-, three-... ladder formation of the
dual parton (DPM) \cite{cstt} or quark-gluon string (QGSM) \cite{kaid, kter, 
mwal} models, and/or to different 
(soft, hard) types of interactions \cite{ghal, gugo}. They become 
increasingly important as collision energy is increased.
These subprocesses are related to the matter state during the collision
(e.g., there are speculations about nonhomogeneous matter distribution in
impact parameters \cite{bour}, not to speak of quark-gluon plasma 
\cite{mcle} behaving as a liquid \cite{shur} etc).

Theoretical description of the subprocesses differs drastically in
$e^+e^-$ and $p\bar p$ processes. Jet evolution in $e^+e^-$ is well described
by the perturbative QCD equations with the only adjustable parameter
$\Lambda _{QCD}$. This parameter is approximately known from other
characteristics and is, therefore, bounded. The production of perturbative
gluons and quark-antiquark pairs can be described in terms of dipole or
"antenna" radiation. Color interference plays a crucial role. Many predictions 
of the perturbative QCD approach have been confirmed by experimental data.
What concerns multiplicity distributions, their shape in $e^+e^-$ processes
is known only implicitly from studies of their moments. This is determined by 
the fact that the QCD evolution equations are formulated in terms of the 
generating function. They can be rewritten as equations for the moments but
not directly for probabilities of $n$ parton emission. Their solutions up to
higher order perturbative QCD \cite{13} have predicted some completely new 
features of the moments. The moments contain also complete information about 
the distribution. However, the reconstruction of the shape of the distribution
from them is not a trivial task.
Some attempts to solve the inverse problem \cite{dkmt, yuri} and 
get directly the shape of the multiplicity distribution were successful only 
in the lowest order perturbative QCD approximations with several additional 
assumptions.

The situation with hadronic processes is, in some respect, more complicated. 
The confinement property is essential, and perturbative QCD methods can not
be directly applied. Therefore, some models have been developed.
Hadron interactions used to be considered as proceeding via collisions of
their constituent partons. In pre-parton times, their role was played by pions, 
and one-meson exchange model \cite{dche} dominated. Pions were treated as 
hadron constituents. Their high energy interaction produced a ladder of 
one-pion $t$-channel exchanges with blobs of low energy pion-pion 
interactions. This is the content 
of the multiperipheral model. These blobs were first interpreted as 
$\rho $-mesons \cite{afst} and later called fireballs \cite{acdr}, clusters 
\cite{ddun} or clans \cite{gvh} when higher mass objects were considered. 
Multiperipheral dynamics tells us that the number of these blobs is distributed 
according to the Poisson law. It was argued that its convolution with the 
distribution of the number of pions produced in each center can lead to the 
negative binomial distribution (NBD) of created particles. This supposition 
fits experimental data on multiplicity distributions of $pp$ and $p\bar p$ 
reactions at tens of GeV quite well. However, at higher energies this fit by 
a single NBD becomes unsatisfactory. A shoulder appears at high multiplicities. 
Sums of NBD with different parameters were used \cite{gugo} to get 
agreement with experiment. Better fits are achieved at 
the expense of a larger number of adjustable parameters.
These shortcomings can be minimized if one assumes that each high energy binary 
parton collision is independent of some others simultaneously proceeding.
With this supposition, the whole process is described as a set of 
independent pair parton interactions (IPPI-model, proposed in 
\cite{ippi}). 
Effective energy of a pair of partons does not depend on how many other
pairs interact and what are these interacting partons (quarks or gluons). 
The number of adjustable parameters does not increase compared to a 
single NBD if the probabilities of $j$ pairs interactions and the number of 
active pairs are known.

Earlier, a somewhat different way to account for multiple parton collisions, 
which generalized the multiperipheral one-ladder model in the framework
of the eikonal approximation, was proposed in DPM 
and QGSM (the latter one takes into account the reggeization of exchanged
particles). They differ from IPPI by probabilities of processes with
different number of active parton pairs and by multiplicity distributions
of final particles. Also, the Lund model \cite{ande} with its parton cascades 
and the string hadronization due to linear increasing QCD potential has been 
extremely successful in describing many features of multiparticle production. 
Several Monte Carlo programs implement this model to provide some hints
to experimentalists at present day experiments and give predictions at even
higher energies (like PYTHIA, FRITIOF etc.). Some assumptions have to be used
for the hadronization of partons at the final stage (e.g., these assumptions 
differ in PYTHIA and HERWIG cluster model). More important, the predictions at 
higher energies (in particular, for multiplicity distributions) also differ in 
these models, and it is necessary to try various approaches and confront 
them to experiment when LHC enters the operation.

\section{Moments of multiplicity distributions.}

The shape of multiplicity distributions is so complicated that it is 
difficult to get any analytical expression for it from the solution
of QCD equations. It has been obtained only in the simplest 
perturbative approximations \cite{dkmt, yuri}. In particular, it 
has been demonstrated \cite{yuri} that the recoil has a profound 
effect on the multiplicity distribution in QCD jets. It becomes 
much narrower than according to the leading perturbative term, where
energy conservation is not taken into account. However, 
no shoulders appear. On the contrary, the tail of the distribution
is stronger suppressed.

The alternative and, in some sense, more accurate approach is proposed 
by studies of moments of the distribution. One can get some QCD 
predictions for these moments \cite{13} up to the higher order of the
perturbative expansion. The moments of various ranks contain complete 
information about multiplicity distributions. Hence, the shape and energy 
evolution of multiplicity distributions can be quantitatively described by 
the rank dependence and energy behavior of their moments. Moment analysis of
multiplicity distributions can be also performed for the models of 
hadronic processes as well as for experimental data. Therefore, this 
approach is common for all processes and methods of analysis.

To introduce moments on most general grounds, we write the generating 
function $G(E,z)$ of the multiplicity distribution $P(n, E)$ 
\be
G(E,z) = \sum_{n=0}^{\infty }P(n,E)(1+z)^{n}.              \label{3}
\ee
The multiplicity distribution is obtained from the generating function as
\be
P(n)=\frac {1}{n!}\frac {d^{n}G(E,z)}{dz^{n}}\vline _{z=-1}.  \label{pne} 
\ee
In what follows, we will use the so-called unnormalized factorial ${\cal F}_q$
and cumulant ${\cal K}_q$ moments defined according to the formulas
\be
{\cal F}_{q} = \sum_{n} P(n)n(n-1)...(n-q+1) =
 \frac {d^{q}G(E,z)}{dz^{q}}\vline _{z=0}, 
\label{4}
\ee
\be
{\cal K}_{q} = \frac {d^{q}\ln G(E,z)}{dz^{q}}\vline _{z=0}. \label{5}
\ee
They determine correspondingly the total and genuine, i.e., irreducible
to lower order correlations among the 
particles produced (for more details see \cite{ddki, dgar}).
First factorial moment defines the average multiplicity ${\cal F}_{1} =
\langle n\rangle$, second one is related to the width (dispersion) of the 
multiplicity distribution ${\cal F}_{2} =\langle n(n-1)\rangle$, etc.
Factorial and cumulant moments are not independent. They are related by
the formula
\be
{\cal F}_{q} =\sum_{m=0}^{q-1}C_{q-1}^m{\cal K}_{q-m}{\cal F}_{m}, \label{fkf}
\ee
where
\be
C^m_{q-1}=\frac {(q-1)!}{m!(q-m-1)!}       \label{cmq}
\ee
are the binomial coefficients.

Since both ${\cal F}_q$ and ${\cal K}_q$ strongly increase with
their rank and energy, the ratio
\be
H_q={\cal K}_q/{\cal F}_q,  \label{hq}
\ee
first introduced in \cite{13}, is especially useful due to partial cancellation 
of these dependences. More important is that some valuable predictions 
about its behavior can be obtained in perturbative QCD. Also, it will be shown 
below that $H_q$ moments of the IPPI-model depend on smaller number of its 
adjustable parameters than factorial and cumulant moments. Thus, even though
${\cal F}_q$, ${\cal K}_q$ and $H_q$ are interrelated, they can provide
knowledge about different facets of the same multiplicity distribution. 

It is easy to find the ratio $H_q$ from iterative formulas
\be
H_q=1-\sum_{p=1}^{q-1}\frac{\Gamma(q)}{\Gamma(p+1)\Gamma(q-p)}H_{q-p}\frac
{{\cal F}_p{\cal F}_{q-p}}{{\cal F}_q},   \label{hqfq}
\ee
once the factorial moments have been evaluated. 

The factorial moments ${\cal F}_q$'s 
are always positive by definition (Eq. (\ref{4})). The cumulant moments 
${\cal K}_q$'s can change sign. They are equal to 0 for Poisson 
distribution. Consequently, $H_q= 0$ in this case.

Let us emphasize that
$H_q$ moments are very sensitive to minute details of multiplicity distributions
and can be used to distinguish between different models and experimental data.

\section{QCD on moments in $e^+e^-$ collisions.}

All moments can be calculated from the generating function as explained above.
The generating functions for quark and gluon jets satisfy definite equations
in perturbative QCD (see \cite{dkmt, dgar}). They are
\begin{eqnarray}
&G_{G}^{\prime }&= \int_{0}^{1}dxK_{G}^{G}(x)\gamma _{0}^{2}[G_{G}(y+\ln x)G_{G}
(y+\ln (1-x)) - G_{G}(y)] \nonumber \\ 
&+&n_{f}\int _{0}^{1}dxK_{G}^{F}(x)\gamma _{0}^{2}
[G_{F}(y+\ln x)G_{F}(y+\ln (1-x)) - G_{G}(y)] ,   \label{50}
\end{eqnarray}
\begin{equation}
G_{F}^{\prime } = \int _{0}^{1}dxK_{F}^{G}(x)\gamma _{0}^{2}[G_{G}(y+\ln x)
G_{F}(y+\ln (1-x)) - G_{F}(y)] ,                                   \label{51}
\end{equation}
where the labels $G$ and $F$ correspond to gluons and quarks,
the energy scale of the process is defined by
$y=\ln Q/Q_{0}, \; Q=p\Theta$ is a virtuality of a jet, $p\approx \sqrt s /2$ 
is its momentum, $\; \Theta $ opening angle, $\; Q_{0}$=const.
Here $G^{\prime }(y)=dG/dy ,$ $ n_f$ is the number of active flavors,
\begin{equation}
\gamma _{0}^{2} =\frac {2N_{c}\alpha _S}{\pi } .               \label{52}
\end{equation}
The running coupling constant in the two-loop approximation is
\begin{equation}
\alpha _{S}(y)=\frac {2\pi }{\beta _{0}y}\left( 1-\frac {\beta _1}
{\beta _{0}^{2}}\cdot \frac {\ln 2y}{y}\right)+O(y^{-3}), \label{al}
\end{equation}
where
\begin{equation}
 \beta _{0}=\frac {11N_{c}-2n_f}{3}, \;\;\;\;\;\;
 \beta _1 =\frac {17N_c^2-n_f(5N_c+3C_F)}{3}.
 \label{be}
\end{equation}
The kernels of the equations are
\begin{equation}
K_{G}^{G}(x) = \frac {1}{x} - (1-x)[2-x(1-x)] ,    \label{53}
\end{equation}
\begin{equation}
K_{G}^{F}(x) = \frac {1}{4N_c}[x^{2}+(1-x)^{2}] ,  \label{54}
\end{equation}
\begin{equation}
K_{F}^{G}(x) = \frac {C_F}{N_c}\left[ \frac {1}{x}-1+\frac {x}{2}\right] ,   
\label{55}
\end{equation}
$N_c$=3 is the number of colors, and $C_{F}=(N_{c}^{2}-1)/2N_{c}
=4/3$ in QCD.  

Herefrom, one can get equations for any moment of the multiplicity distribution
both for quark and gluon jets. One should just equate the terms with the same
powers of $u=1+z$ in both sides of the equations where expressions (\ref{3})
are substituted for both quarks and gluons.

In particular, the non-trivial energy dependence of mean multiplicity in quark 
and gluon jets has been predicted. Within two lowest perturbative QCD 
approximations it has a common behavior
\begin{equation}
\langle n_{G,F}\rangle=A_{G,F}y^{-a_{1}c^2 }\exp ( 2c\sqrt y),   \label{nm}                        \label{mul}
\end{equation}
where $A_{G,F}$=const, $\; c=(4N_c/\beta _0)^{1/2}, \;  a_1\approx 0.3 $.

Main features of the solutions can be demonstrated in gluodynamics where
only first equation with $n_f=0$ is considered. At asymptotically high 
energies it can be reduced \cite{dkmt} to the differential equation
\be
[\ln G(y)]''=\gamma_0^2[G(y)-1].    \label{gas}
\ee
From the moments definitions, one can easily guess that this equation determines 
the asymptotic behavior of $H_q$ because $\ln G$ in the left-hand side gives
rise to ${\cal K}_{q}$ and $G$ in the right-hand side to ${\cal F}_{q}$. 
The second derivative in the left-hand side would result in the factor $q^2$.
Thus, it can be shown \cite{13} that asymptotical ($y\rightarrow \infty $)
values of $H_q$ moments are positive and decrease as $q^{-2}$. At present
energies they become negative at some values of $q$ and reveal the negative 
minimum at
\be
q_{min}=\frac {1}{h_1\gamma_0}+0.5+O(\gamma _0),  \label{qmin}
\ee
where $h_1=b/8N_c=11/24, \;\; b=11N_c/3-2n_f/3$. At $Z^0$ energy 
$\alpha _S\approx 0.12$, and this minimum is located at about $q\approx 5$. 
It moves to 
higher ranks with energy increase because the coupling strength decreases. Some 
hints to possible oscillations of $H_q$ vs $q$ at higher ranks at LEP energies 
were obtained in \cite{13}. They were obtained with account of recoil effects
in the higher order perturbative QCD and stressed the importance of energy
conservation in high energy processes so often mentioned by Bo Andersson
(see \cite{ande}). The approximate solution of the gluodynamics 
equation for the generating function \cite{41} agrees with this conclusion 
and predicts the oscillating behavior at higher ranks. 
The same conclusions were obtained from exact solution of equations for
quark and gluon jets in the framework of fixed coupling QCD \cite{21}.
A recent exact numerical solution of the gluodynamics equation in a wide 
energy interval \cite{bfoc} coincides with the qualitative features of
multiplicity distributions described above. These oscillations were confirmed
by experimental data for $e^+e^-$ collisions and found also for 
hadron-initiated processes 
first in \cite{dabg}, later in \cite{sld} and most recently in \cite{l3}.
This will be demonstrated below in Fig. 10.

However, one should be warned that the amplitudes of oscillations strongly
depend on a multiplicity distribution cut-off due to limited experimental
statistics (or by another reasoning) if it is done at rather low multiplicities
\cite{dcut}. 
Usually there are no such cut-offs in analytical expressions for $H_q$. 
One can control the influence of cut-offs by shifting them appropriately. 
The qualitative features of $H_q$ behavior persist nevertheless. In what 
follows, we consider
very high energy processes where the cut-off due to experimental statistics
is practically insignificant. Numerical estimates of the relation between
maximum multiplicity measured and effective ranks of the moments is 
given in Appendix.

\section{Negative binomial distribution and IPPI-model.}

Independently of progress in perturbative QCD calculations, the NBD-fits 
of multiplicity distributions were attempted both for $e^+e^-$ and $pp 
(p\bar p)$ collisions \cite{gugo, gug1, sark}. The single 
NBD-parameterization is
\be
P_{NBD}(n,E)=\frac{\Gamma(n+k)}{\Gamma(n+1)\Gamma(k)}\left (\frac{m}{k}\right)^n
\left(1+\frac{m}{k}\right)^{-n-k},  \label{pnbd}
\ee
where $\Gamma $ denotes the gamma-function. This distribution has two 
adjustable parameters $m(E)$ and $k(E)$ which depend on energy.
However the simple fit by the formula (\ref{pnbd}) is valid till the 
shoulders appear. In that case, this formula is often replaced \cite{gugo}
by the hybrid NBD which simply sums up two or more expressions 
like (\ref{pnbd}). Each of them has its own parameters $m_j, k_j$.
These distributions are weighted with the energy dependent probability 
factors $w_j$ which sum up to 1. Correspondingly, the number of adjustable 
parameters drastically increases if the distributions are completely unrelated. 

It was proposed recently \cite{ippi} that hadron interactions can be 
represented by a set of Independent Pair Parton Interactions (IPPI-model).
This means that colliding hadrons are considered as "clouds" of partons
which interact pairwise. It is assumed that each binary parton collision 
is described by the same NBD distribution. The only justification for this
supposition is given by previous fits of multiplicity distributions at lower
energies. Then the convolution of these distributions, subject to a condition 
that the sum of binary collision multiplicities is the total multiplicity $n$, 
leads \cite{ippi} to a common distribution 
\be
P(n; m, k)=\sum_{j=1}^{j_{max}}w_jP_{NBD}(n; jm, jk).   \label{pnj}
\ee
This is the main equation of
IPPI-model. One gets a sum of negative binomial distributions with 
shifted maxima and larger widths for a larger number of collisions. 
No new adjustable parameters appear in the distribution 
for $j$ pairs of colliding partons. The probabilities $w_j$ 
are determined by collision dynamics and, in principle, can be 
evaluated if some model is adopted (e.g., see \cite{kter, mwal}).
Independence of parton pairs interactions implies that at very high energies 
$w_j$ is a product of $j$ probabilities $w_1$ for one pair.
Then from the normalization condition 
\be
\sum_{j=1}^{j_{max}}w_j=\sum_{j=1}^{j_{max}}w^j_1=1   \label{w1w1}
\ee
one can find $w_1$ if $j_{max}$, which is determined by the maximum 
number of parton interactions at a given energy, is known. This is
the only new parameter. It depends on energy. Thus three parameters
are sufficient to describe multiplicity distributions at any energy.
Moreover, asymptotically $j_{max}\rightarrow \infty$ and $w_1=0.5$
according to (\ref{w1w1}). 

The factorial moments of the distribution (\ref{pnj}) are
\be
{\cal F}_{q} = \sum_{j=1}^{j_{max}}w_j\frac{\Gamma(jk+q)}{\Gamma(jk)}
\left(\frac{m}{k}\right)^q=f_q(k)\left(\frac{m}{k}\right)^q   \label{fphi}
\ee
with
\be
f_q(k)=\sum_{j=1}^{j_{max}}w_j\frac{\Gamma(jk+q)}{\Gamma(jk)}. \label{phiq}
\ee
For $H_q$ moments one gets
\be
H_q=1-\sum_{p=1}^{q-1}\frac{\Gamma(q)}{\Gamma(p+1)\Gamma(q-p)}H_{q-p}\frac
{f_pf_{q-p}}{f_q}.   \label{hqph}
\ee
Note that according to Eqs (\ref{phiq}), (\ref{hqph}) $H_q$ are functions 
of the parameter $k$ only and do not depend on $m$ in IPPI-model.
This remarkable property of $H_q$ moments provides an opportunity to
fit them by one parameter. It is nontrivial because the oscillating shapes
of $H_q$ are quite complicated as shown below.

Once the parameter $k$ is found from fits of $H_q$, it is possible to get
another parameter $m$ rewriting Eq. (\ref{fphi}) as follows
\be
m=k\left(\frac {{\cal F}_{q}}{f_q(k)}\right)^{1/q}. \label{mkfp}
\ee
This formula is a sensitive test for the whole approach because it states
that the definite ratio of $q$-dependent functions to the power $1/q$ becomes
$q$-independent if the model is correct. Moreover, this statement should be 
valid only for those values of $k$ which are determined from $H_q$ fits.
Therefore, it can be considered as a criterion of a proper choice of $k$ 
and of the model validity, in general. This criterion of constancy of $m$
happens to be extremely sensitive to the choice of $k$ as shown below.
 
In the paper \cite{mwal}, the energy dependence of the probabilities $w_j$ 
was estimated according to the multiladder exchange model \cite{kter}
(its various modifications are known as DPM - Dual Parton Model or QGSM -
Quark-Gluon String Model). 
They are given by the following normalized expressions
\be
w_j(\xi_j)=\frac{p_j}{\sum_{j=1}^{j_{max}}p_j}
=\frac {1}{jZ_j(\sum_{j=1}^{j_{max}}p_j)}
\left( 1-e^{-Z_j}\sum_{i=0}^{j-1}\frac {Z_j^i}{i!}\right)      \label{kter}
\ee
where 
\be
\xi_j=\ln(s/s_0j^2), \;\;  Z_j=\frac {2C\gamma}{R^2+\alpha_P'\xi_j}
\left(\frac{s}{s_0j^2}\right)^{\Delta}  \label{xiz}
\ee
with numerical parameters obtained from fits of
experimental data on total and elastic scattering cross sections: 
$\gamma=3.64$ GeV$^{-2}$, $\,$ $R^2=3.56$ GeV$^{-2}$,
$\,$ $C=1.5,$ $\,$ $\Delta=\alpha_P-1=0.08,$ $\,$ $\alpha_P'=0.25$ GeV$^{-2}$,
$s_0$ =1 GeV$^2$. It is seen that each $w_j$ depends on 6 adjustable 
parameters in these models.

Below, we will use both possibilities (\ref{w1w1}) and (\ref{kter}) 
in our attempts to describe experimental data.
The probabilities $w_j$ are different for them (see Table 1).
In IPPI-model they decrease exponentially with the number of active partons,
while in the ladder models they are inverse proportional to this number
with additional suppression at large $j$ due to the term in brackets in
(\ref{kter}). This is the result of the modified eikonal approximation.
Let us stress that, when we use expressions (\ref{kter}) for probabilities,
this does not directly imply comparison with DPM-QGSM because in our case 
NBD is chosen for the multiplicity distribution in a single ladder, while 
it is Poisson distribution for resonances in DPM-QGSM. Thus, we will call it 
the modified ladder model.

We show the values $w_j$ for $j_{max}$=3 -- 6 pairs calculated 
according to Eq. (\ref{w1w1}) in the left-hand side of Table 1 and according
to Eq. (\ref{kter}) in its right-hand side. These values of
$j_{max}$ are chosen because previous analysis of experimental data \cite{walk}
has shown that 2 pairs become active at energy about 120 GeV and the number
of binary collisions increases with energy increasing. Thus, namely these 
numbers will be used in comparison with experiment at higher energies.
In particular, we shall choose $j_{max}=3$ at 300 and 546 GeV, 4 at 1000 and
1800 GeV, 5 at 14 TeV and 6 at 100 TeV (see below).\\

{\bf Table 1.  }\\

The values of $w_j$ according to (\ref{w1w1}) (left-hand side) and
(\ref{kter}) (right-hand side).\\

\begin{tabular}{|c||c|c|c|c||c|c|c|c|} \hline
$j_{max}$& 3&  4&     5&     6   &    3 &    4 &    5 &     6  \\ \hline
$w_1$& 0.544& 0.519& 0.509& 0.504& 0.562& 0.501& 0.450& 0.410\\ 
$w_2$& 0.295& 0.269& 0.259& 0.254& 0.278& 0.255& 0.236& 0.219\\
$w_3$& 0.161& 0.140& 0.131& 0.128& 0.160& 0.153& 0.152& 0.147\\
$w_4$& 0    & 0.072& 0.067& 0.065& 0    & 0.091& 0.100& 0.104\\ 
$w_5$& 0    & 0    & 0.034& 0.033& 0    & 0    & 0.062& 0.073\\
$w_6$& 0    & 0    & 0    & 0.016& 0    & 0    & 0    & 0.047\\ \hline
\end{tabular}                    \\

One can clearly see the difference between the two approaches. The value of $w_1$
is always larger than 0.5 in the IPPI-model, while it can become less than 0.5 
in the (modified) ladder model \cite{cstt, kter} at high energies. 
In the ladder model, $w_j$ depend explicitely on energy (not only on $j_{max}$ 
cut-off). We show their 
values at 546 and 1800 GeV in the right-hand side columns of $j_{max}$=3 
and 4. Those at 300 and 1000 GeV are larger for $w_1$ by about 1$\%$ and smaller 
for $w_3$ by about 3$\%$.  When energy increases, the processes with 
a larger number of active pairs play more important role in the modified 
ladder approach compared to IPPI-model. Thus, the $j_{max}$ cut-off is 
also more essential there.

In principle, one can immediately try a two-parameter fit of experimental
multiplicity distributions using Eq. (\ref{pnj}) if $w_j$ are known
according to Eqs. (\ref{w1w1}) or (\ref{kter}).
However, the use of their moments is preferred.

\section{Comparison with experiment.}

Let us begin with hadronic collisions and then compare them with other 
processes.

We have compared \cite{ippi} IPPI-model conclusions with experimental 
(but extrapolated \cite{walk, arn} to the full phase space)
multiplicity distributions of E735 \cite{e73, e735} collaboration for 
$p\bar p$ collisions at Tevatron energies 300, 546, 1000 and 1800 GeV.
The multiplicity of charged particles was divided by 2 to get the multiplicity
of particles with the same charge. Then the above formulas for moments were
used. Correspondingly, the parameters $m$ and $k$ refer to these distributions.
The parameter $j_{max}$ is chosen according to prescriptions discussed above.

Factorial and $H_q$ moments were obtained from experimental data on 
$P(n)$ according to Eqs. (\ref{4}), (\ref{hqfq}).
Experimental $H_q$ moments were fitted by Eq. (\ref{hqph}) 
to get the parameters $k(E)$. We show in Fig. 1 how perfect are these fits
at 1.8 TeV for $k$ equal to 3.7 (solid line) and 4.4 (dash-dotted line). 
At this energy, we considered
four active parton pairs with $w_j$ given by Eq. (\ref{w1w1}). It is
surprising that oscillations of $H_q$ moments are well reproduced with
one adjustable parameter $k$. The general tendency of this quite 
complicated oscillatory dependence is clearly seen.

With these values of the parameter $k$, we have checked
whether $m$ is constant as a function of $q$ as required by Eq. (\ref{mkfp}).
The $m(q)$ dependence is shown in Fig. 2 for the same values of $k$ and for
much larger value 7.5. The constancy of $m$ is fulfilled with an accuracy
better than 1$\%$ for $k=4.4$ up to $q=16$. The upper and lower lines 
in Fig. 2 demonstrate clearly that this condition bounds substantially 
admissible variations of $k$. 

The same-charge multiplicity distribution at 1.8 TeV has been fitted with 
parameters $m=12.94$ and $k$=4.4 as shown in Fig. 3 (solid line). 
To estimate the accuracy of the fit, we calculated 
$\sum_{n=1}^{125}(P_t(n)-P_e(n))^2/\Delta ^2$ over all 125 experimental points. 
Here, $P_t, P_e$ are theoretical and experimental distributions and $\Delta$ is
the total experimental error. It includes both statistical and systematical
errors. Note that the latter ones are large at low multiplicities in E735 data.
This sum is equal to 50 for 125 degrees of freedom. No minimization of it was 
attempted. This is twice better than the three-parameter fit by the 
generalized NBD considered in \cite{heg}.
Poisson distribution of particles in binary collisions is completely 
excluded. This is shown in Fig. 3 by the dash-dotted line.

The same procedure has been applied to data at energies 300, 546, 1000 GeV. 
As stated above, we have assumed that 3 binary parton collisions are 
active at 300 and 546 GeV and 4 at 1000 GeV. We 
plot in Figs. 5 and 6 the energy dependence of parameters $m$ and $k$. 
The parameter $m$ increases with energy logarithmically (Fig. 5). This is 
expected because increase of $M_1=\sum w_jj$ due to increasing number of 
active pairs at these energies leads to somewhat faster 
than logarithmical increase of average multiplicity in accordance with
experimental observations. The energy
dependence of $k$ is more complicated and rather irregular (Fig. 6). 

We tried to ascribe the latter to the fact that the effective values of $k$, 
which we actually find from these fits, depend on the effective number of 
parton interactions, i.e. on $w_j$ variation at a threshold.
The threshold effects can be important in this energy region. Then, the simple 
relation (\ref{w1w1}) is invalid. This influences the functions $f_q(k)$
(\ref{phiq}) and, consequently, $H_q$ calculated from Eq. (\ref{hqph}).
One can reduce the effective number of active pairs to about 2.5 at 300 GeV 
and 3.5 at 1000 GeV if chooses the following values of $w_j$: 0.59,$\,$ 0.34,
$\,$ 0.07 at 300 GeV and 0.54,$\,$ 0.29,$\,$ 0.14,$\,$ 0.03 at 1000 TeV instead
of those calculated according to (\ref{w1w1}) and shown in Table 1. This 
gives rise to values of $k$ which are not drastically different from previous
ones. However, the quality of fits becomes worse.
Fits with 2 active pairs at 300 GeV and 3 pairs at 1000 GeV fail completely. 

Hence, we have to conclude that this effect results from dynamics of hadron
interactions which is not understood yet and should be incorporated in
the model. The preliminary explanation of this effect could be that at
the thresholds of a new pair formation the previous active pairs produce
more squeezed multiplicity distributions due to smaller phase-space room 
available for them because of a newcomer. Therefore, the single pair 
dispersion would decrease and the $k$ values increase. It would imply 
that thresholds are marked not only by the change of $w_j$ shown in Table 1
but also by the variation of the parameter $k$.

The threshold effects become less important at higher energies. We assume 
that there are 5 active pairs at 14 TeV and 6 at 100 TeV. Then we extrapolate 
to these energies. The parameter $m$ becomes equal to 19.2 at 14 TeV and 25.2 
at 100 TeV if logarithmical dependence is adopted as shown in Fig. 5 by the
straight line.
We choose two values of $k$ equal to 4.4 and 8 since we do not know
which one is responsible for thresholds. The predicted 
multiplicity distributions with these parameters are plotted in Fig. 7.
The oscillations of $H_q$ still persist at these energies (see Fig. 8).
The minima are however shifted to $q=6$ at 14 TeV and 7 at 100 TeV as expected.

The fit at 1.8 TeV with an approximation of $w_j$ according to the 
modified ladder model (\ref{kter}) with NBD for a binary parton collision
is almost as successful as 
the fit with values of $w_j$ given by IPPI-model. However, some difference at
14 TeV (see Fig. 7) and especially at 100 TeV between these models is predicted.
To keep the same mean multiplicity in both models at the same energy, we have 
chosen different values of $m$ according to $\langle n \rangle = m\sum w_jj$ 
and $w_j$ values shown in Table 1. Namely, their ratios are $m_{IPPI}/m_{lad}$
=0.988, $\,$ 1.039, $\,$ 1.145, $\,$ 1.227 for $j_{max}$=3, $\,$ 4, $\,$ 5, 
$\,$ 6, correspondingly. This shows that the maximum of the distribution 
moves to smaller multiplicities and its width becomes larger in the modified 
ladder model compared to IPPI-model with energy increasing.

Surely, one should not overestimate the success of the IPPI-model in its present
initial state. It has been applied just to multiplicity distributions. 
For more detailed properties, say, rapidity distributions, one would
need a model for the corresponding features of the one-pair process.
Moreover, the screening effect (often described by the triple Pomeron vertex)
will probably become more important at higher energies. All these features
are somehow implemented in the well known Monte Carlo programs PYTHIA
\cite{pyth}, HERWIG \cite{herw} and DPM-QGSM \cite{cstt, kter}
. However, for the latter one, e.g., the multiplicity
distribution for a single ladder is given by the Poisson distribution of 
emission centers (resonances) convoluted with their decay properties,
and probabilities $w_j$ contain several adjustable parameters. It differs
from the IPPI-model. The latter 
approach proposes more economic way with a smaller number of such parameters.
What concerns the further development of the event generator codes, it is 
tempting to incorporate there the above approach with a negative binomial
distribution of particles created by a single parton pair, and confront the 
results to a wider set of experimental data. We intend to do it
later to learn how it influences other characteristics.

\section{Is hadronic production similar in various processes?}

This question was first raised by the statement of \cite{basi} that the average
multiplicities in $e^+e^-$ and $pp (p\bar p)$ processes increase with energy 
in a similar way. Recently, PHOBOS Collaboration \cite{rhic} claimed even
that the energy behavior of mean multiplicities in all processes is similar.
Therefrom, it has been concluded that the dynamics of all hadronic processes 
is the same.
Beside our general belief in QCD, we can not claim that other characteristics
of multiple production processes initiated by different partners coincide.
To answer the above question, we compare characteristics of multiplicity
distributions for processes initiated by different partners.

{\bf Average multiplicities.}
Total yields of charged particles in high energy $e^+e^-$, $p\bar p$, $pp$ and 
central AA collisions become similar if special rescaling has been done. The 
average charged particle multiplicity in $pp/p\bar p$ is similar \cite{basi} to that 
for $e^+e^-$ collisions if the effective energy $s_{eff}$ is inserted in place
of $s$, where $\sqrt{s_{eff}}$ is the $pp/p\bar p$ center-of-mass energy minus 
the energy of the leading particles. In practice, $\sqrt{s_{eff}}=\sqrt s/2$ is
chosen. This corresponds to the horizontal shift of empty squares to the
diamond positions in Fig. 9 borrowed from \cite{rhic}. Then the diamonds lie
very close to the dashed line, which shows QCD predictions for multiplicities 
in $e^+e^-$ collisions according to (\ref{nm}).
For central nucleus-nucleus collisions the particle yields have been scaled by 
the number of participating nucleon pairs $N_{part}/2$. Then the energy 
dependence of mean multiplicities is very similar for all these processes at 
energies exceeding 10 GeV up to 1 TeV, and even at lower energies for $e^+e^-$
compared with $pp/p\bar p$. This is well demonstrated in Fig. 9. 

However, the situation changes at higher energies. In Table 2 we show 
experimentally measured mean multiplicities at Tevatron energies (first row). 
They are compared with the results of IPPI-model in the second row, where the
IPPI predictions at 14 TeV (LHC) and 100 TeV are also shown. According to the
above hypothesis, these values should coincide with QCD predictions 
(\ref{nm}) at twice
smaller energy. The latter ones are presented in the third row. The difference 
between them and experimentally measured values of the first row is
demonstrated in the fourth row for Tevatron energies, while at higher energies
the difference of QCD and IPPI predictions is shown.\\

{\bf Table 2.  }\\

The mean multiplicities at Tevatron and higher energies.\\

\begin{tabular}{|c||c|c|c|c||c|c|} \hline
$\sqrt s,$ GeV& 300&  546&$ 10^3$&1.8$\cdot 10^3$&1.4$\cdot 10^4$&$10^5$  \\ \hline
$n_{p\bar p, exper.}$& 25.4& 30.5& 39.5& 45.8& -- & -- \\ 
$n_{p\bar p, theor.}$& 24.1& 30.0& 39.4& 45.7& 71.6& 97.0\\ \hline
$n_{e^+e^-, theor.}$& 25.3& 31.8& 39.9& 49.2& 98.2& 180\\
$n_{e^+e^-}-n_{p\bar p}$& 0.1 & 1.3& 0.4& 3.4& 26.6& 83\\  \hline
\end{tabular}                    \\

It is seen that both experimental and theoretical values of average
multiplicity coincide pretty well in $p\bar p$ and $e^+e^-$ up to 1 TeV.
However, already at Tevatron energy 1.8 TeV the rescaled $p\bar p$
multiplicity is lower by 3.4 charged particles. This difference between
rescaled predictions of IPPI-model for $p\bar p$ and QCD for $e^+e^-$ becomes
extremely large at LHC (26.6), and even more so at higher energies.

With these observations, one tends to claim (if at all!)
the approximate quantitative similarity of all processes 
up to 1 TeV and "a universal mechanism of particle production 
in strongly-interacting systems controlled mainly by the amount of energy
available for particle production" \cite{rhic} only at energies below
highest Tevatron values. The situation becomes even worse if we compare other
features of multiplicity distributions.

{\bf $H_q$ moments.}
First, we have calculated \cite{dnbs} $H_q$ moments for experimental 
multiplicity
distributions in various high energy processes. They are shown in Fig. 10.
These moments weakly depend on energy in the energy regions available to
present experiments. Their oscillating shape is typical for all processes.
However, one notices that amplitudes of $H_q$ oscillations in Fig. 10
differ in various reactions. They are larger
for processes with participants possessing more complicated internal structure.
For example, amplitudes in $e^+e^-$ are about two orders of magnitude smaller
than those in $pp$. Both QCD applied to $e^+e^-$ \cite{dgar, bfoc} and models 
of $pp/p\bar p$ \cite{ippi} can fit these observations.

There is, however, one definite QCD prediction, which allows to ask a question 
whether QCD and, e.g., IPPI-model are compatible, in principle. This is
the asymptotical behavior of $H_q$ moments in QCD. They should behave
\cite{13} as $H_q^{as}=1/q^2$. One can also determine asymptotics of
$H_q$ moments in the IPPI-model and compare both approaches \cite{dmo}. 
These values are noticeably larger than QCD predictions of $1/q^2$. 
Thus, QCD and IPPI-model have different asymptotics. 
It is an open question whether other asymptotic relations for probabilities
of multiparton interactions different from those adopted in IPPI-model can 
be found which would lead to the same behavior of $H_q$ moments in $p\bar p$ 
and $e^+e^-$ collisions. Only then one can hope to declare for analogy 
between these processes.

{\bf Fractal properties.}
The particle density within the phase space in individual events is much more
structured and irregular than in the sample averaged distribution. However, 
even for the latter ones fluctuations depend on the phase space of a sample.
The smaller is the phase space the larger are the fluctuations. This can be 
described by the behavior of normalized factorial moments 
$F_q={\cal F}_{q}/\langle n\rangle ^q$ as functions of the amount of phase
space available. In one-dimensional case of the rapidity distributions within
the interval $\delta y$, the power-like behavior $F_q(\delta y)\propto
(\delta y)^{-\phi (q)}$ for $\delta y\rightarrow 0$  and $\phi (q)>0$
would correspond to intermittent phenomenon well known from turbulence.
This also shows that created particles are distributed in the phase space in
a fractal manner. The anomalous fractal dimension $d_q$ is connected with the
intermittency index $\phi (q)$ by the relation $\phi (q)=(q-1)(d-d_q)$, where
$d$ is the ordinary dimension of a sample studied ($d=1$ for the one-dimensional
rapidity plot). It has been calculated in QCD \cite{owos, ddre, bmpe} only in
the lowest order perturbative approximations. The qualitative features of the
behavior of factorial moments as functions of the bin size observed in 
experiment are well reproduced by QCD.

The anomalous dimensions as functions of the order $q$ derived from 
experimental data are shown in Fig. 11
for various collisions \cite{ddki}. They are quite large for $e^+e^-$, 
become smaller for $hh$ and even more so for $AA$ collisions. This shows
that the more structured are the colliding partners, the stronger smoothed are
the density fluctuations in the phase space. In some way, this observation 
correlates with the enlarged amplitudes of $H_q$ oscillations mentioned above.

In any case, this is another indication that it is premature to claim the
similarity of $e^+e^-$ and hadron initiated processes. We have compared just 
some features of multiplicity distributions. There are many more characteristics
of the processes which can be compared but this is out of the scope of the
present paper.

\section{Conclusions.}

We have briefly described the theoretical approaches to collisions of high energy
particles. They seem to be quite different for various processes. No direct
similarity in multiplicity distributions has been observed. According to
experimental data,
the energy dependence of average multiplicities in different collisions can be 
similar in energy region above 10 GeV if some rescaling procedures are used.
However, this is just the first moment of multiplicity distributions. To speak
about their similarity one should compare other moments. Again, the qualitative 
features of moment oscillations are somewhat similar but quantitatively they
differ. The same can be said about the fractal properties of particle densities
within the phase space. No deep reasoning for corresponding rescaling has been
promoted. Thus, beside our general belief that QCD Lagrangian is at the origin
of all these processes, we can not present any serious arguments in favor of 
similar schemes applicable to the dynamics of the processes. Moreover, QCD 
and considered models of hadron interactions predict different asymptotics 
for some characteristics of $e^+e^-$ and $pp (p\bar p)$.
We have considered here just multiplicity distributions. Other inclusive and 
exclusive characteristics also seem to be different in these processes.

To conclude, multiplicity distributions in various high energy processes
possess some common qualitative features but differ quantitatively.
Theoretical approaches to their description have very little in common.
Thus, it is premature to claim their common dynamical origin independently
of our belief in QCD as a theory of strong interactions.\\

{\bf Acknowledgements.}

This work has been supported in part by the RFBR grants 02-02-16779,
04-02-16445-a, NSH-1936.2003.2.\\

{\bf Appendix.}

The higher is the rank of the moment, the higher multiplicities 
contribute to it. Therefore the high rank moments are extremely 
sensitive to the high multiplicity tail of the distribution. At the same time,
the energy-momentum conservation and experimental 
statistics limitations are important at the tail of the distribution.
Therefore, the question about the limits of applicability of the whole
approach is quite reasonable.

Let us estimate the range of validity of considering large $q$ values of $H_q$
moments imposed by some cut-offs (see also \cite{dcut}).
It is well known that experimental cut-offs of multiplicity distributions 
due to the limited statistics of an experiment can influence the behavior of 
$H_q$ moments. Consequently, they impose some limits on $q$ values allowed to 
be considered when a comparison is done. Higher rank moments can be evaluated
if larger multiplicities have been measured. To estimate the admissible range 
of $q$, we use the results obtained in QCD. Characteristic multiplicities that
determine the moment of the rank $q$ can be found. By inverting this relation,
one can write the asymptotic expression for the characteristic range 
of $q$ \cite{yuri}. This provides the bound $q_{max}\approx
Cn_{max}/\langle n\rangle $ where $C\approx 2.5527$. However, it 
underestimates the factorial moments. Moreover, the first moment is not 
properly normalized (it becomes equal to $2/C$ instead of 1). The 
strongly overestimated values (however, with a correct normalization
of the first moment)
are obtained if $C$ is replaced by 2. Hence, one can say
that the limiting values of $q$ are given by inequalities
\begin{equation}
2n_{max}/\langle n\rangle < q_{max}\leq Cn_{max}/\langle n\rangle.  \label{qmax}
\end{equation}
The ratio $n_{max}/\langle n\rangle $ measured by E735 collaboration at 1.8 TeV
is about 5. Thus, $q_{max}$ should be in the interval between 10 and 13. The 
approximate constancy of $m$ and
proper fits of $H_q$ demonstrated above persist to even higher ranks.\\

{\bf Figure captions.}\\
\begin{tabbing}
Fig. 1.  \= A comparison of $H_q$ moments derived from experimental data at 1.8 TeV \\
         \> (squares) with their values calculated with parameter $k$=4.4\\
         \> (dash-dotted line) and 3.7 (solid line) \cite{ippi}.\\
Fig. 2.  \> The $q$-dependence of $m$ for $k$=4.4 (squares), 3.7 (circles) and 7.5 (triangles) \cite{ippi}.\\
Fig. 3.  \> The multiplicity distribution at 1.8 TeV, its fit at $m=12.94, k=4.4$ (solid line). \\
         \> The dash-dotted line demonstrates what would happen\\
         \> if NBD is replaced by Poisson distribution \cite{ippi}.\\
Fig. 4.  \> The decomposition of the fit in Fig. 3 to 1,$\,$ 2, $\,$ 3 and 4 parton-parton collisions \cite{ippi}.\\
Fig. 5.  \> The energy dependence of $m$ (squares) and its linear extrapolation \\
         \> (circles at 14 and 100 TeV) \cite{ippi}.\\
Fig. 6.  \> The values of $k$ as calculated with $w_j$ satisfying the relation (\ref{w1w1}) \cite{ippi}.\\
Fig. 7.  \> The same-charge multiplicity distributions at 14 TeV and 100 TeV obtained by\\
         \> extrapolation of parameters $m$ and $k$ with 5 active pairs at 14 TeV and 6 at 100 TeV\\
         \> (for IPPI-model: solid line - 14 TeV, $k$=4.4; dash-dotted - 14 TeV, $k$=8; dashed -\\
         \> 100 TeV, $k$=4.4; for the modified ladder model: dotted - 14 TeV, $k$=4.4) \cite{ippi}.\\
Fig. 8.  \> The behavior of $H_q$ predicted at 14 TeV ($k$=4.4 -solid line; $k$=8 - \\
         \> dash-dotted line) \cite{ippi}.\\
Fig. 9.  \> The energy dependence of average multiplicities for various processes (a)\\
         \> and their ratio to QCD prediction for $e^+e^-$ collisions (b) \cite{rhic}.\\
Fig. 10. \> Oscillations of $H_q$ moments for various processes \cite{dnbs}.\\
Fig. 11. \> Anomalous fractal dimensions for various processes \cite{ddki}.   \\
\end{tabbing}

\end{document}